# Three-photon absorption in water-soluble ZnS nanocrystals

Jun He, Wei Ji,[a)] and Jun Mi

Department of Physics, National University of Singapore
2 Science Drive 3, Singapore 117542

Yuangang Zheng and Jackie Y. Ying

Institute of Bioengineering and Nanotechnology
31 Biopolis Way, The Nanos, Singapore 138669

We report on large three-photon absorption (3PA) in glutathione-capped ZnS semiconductor nanocrystals (NCs), determined by both Z-scan and transient transmission techniques with 120-fs laser pulses. The monodispersed, water-soluble ZnS NCs are synthesized by a modified protocol with a mean diameter of 2.5 nm. Their 3PA cross-section is determined to be ~2.7 × $10^{-78}$ $cm^6$ $s^2$ $photon^{-2}$ at an optimal wavelength of commercial Ti:sapphire femtosecond lasers. This value is nearly one order of magnitude greater than that of CdS NCs, and four to five orders of magnitude higher than those of the previously reported common UV fluorescent dyes.

[a)]Electronic mail: phyjiwei@nus.edu.sg

Semiconductor nanocrystals (NCs) with large multiphoton absorption (MPA) have been the focus of material research for multiphoton fluorescence imaging.[1,2] Compared to common fluorophores, semiconductor NCs have many advantages: broad absorption bands, but very narrow and symmetric emission bands, tunable emission wavelength, longer emission lifetime, and enhanced brightness. While two-photon absorption (2PA) in semiconductor NCs has been widely investigated,[3,4] research effort on their three-photon absorption (3PA) is limited.[5] Although high-quality CdS NCs show strong multiphoton-excited, band-gap emission,[5] the intrinsic toxicity of cadmium places ZnS NCs in an advantageous position. ZnS and related II-VI compounds are also attractive for applications in photonic crystal devices in the visible and near-infrared region due to their high indices of refraction and large band gap, which make them highly transparent in the visible region.[6] Recently, multicolor electroluminescence[7] and fluorescence[8] of doped ZnS NCs have been investigated. Furthermore, Nikesh *et al.* have reported large 2PA in ZnS nanoparticles using picosecond Z-scan technique.[9] Here, we report on the synthesis, characterization, and 3PA measurements of water-soluble ZnS NCs.

The synthesis of ZnS NCs was based on the reaction of zinc chloride and sodium sulfide. The freshly prepared $Na_2S$ solution was added to another solution containing $ZnCl_2$ and glutathione at pH 11.5 with vigorous stirring. The amounts of $ZnCl_2$, $Na_2S$ and GSH were 5, 2 and 6 mmol, respectively, in a total volume of 500 ml. The resulting mixture was heated to 95ºC, and the growth of GSH-capped ZnS NCs took place immediately. The band-gap edge changed from 240 nm to 293 nm in 60 min of heating. The as-prepared NCs with band-gap edge at ~293 nm were precipitated, and washed several times with 2-propanol. The pellet of the NCs was vacuum dried at room temperature overnight, and the final product in the powder form could be redissolved in water. The elemental analysis of the water-soluble ZnS NCs was performed on ELAN 9000/DRC ICP-MS system. Both morphology and size distribution of the NCs were examined with a field emission high-resolution transmission electron microscope (HRTEM) (FEI Tecnai TF-20, 200 kV). The powder X-ray diffraction (XRD) pattern of the vacuum-dried ZnS NCs was obtained with PANalytical X'Pert PRO. The one-photon absorption spectrum was measured on a UV-visible spectrophotometer (Shimadzu, UV-1700). The photoluminescence (PL) and photoluminescence excitation (PLE) spectra were collected with a Jasco FP-6300 spectrofluorometer. They were obtained before and after the pulsed laser irradiation; no measurable difference was observed, showing the high photostability of ZnS NCs in aqueous solution.

Figure 1(a) presents a HRTEM image of the ZnS NCs capped with GSH. The average diameter of the ZnS NCs is 2.5 ± 0.3 nm, which is dependent on the amount of capping agent used in the preparation. In the inset of Fig. 1(a), the higher magnification image reveals the crystalline lattice of the NCs, with a typical lattice spacing of ~ 3.0 Å, which corresponds to the (111) plane of ZnS. A typical XRD pattern of the NCs is shown in Fig. 1(b), which corresponds to the cubic (zinc blende) phase. The XRD analysis[10] indicates that the ZnS NCs have a mean diameter of 2.2 nm, close to the HRTEM result.

The one-photon absorption spectrum of the GSH-capped ZnS NCs in aqueous solution is shown in Fig. 2. A recent theoretical calculation[11] shows that due to the quantum confinement, there is a blue-shift of ~ 450 meV in the bang-gap energy of 2.5-nm-diameter ZnS NCs, in comparison to that of bulk ZnS. Since the band-gap energy for bulk ZnS is ~3.7 eV, the lowest excitonic transition in the NCs, $1S(e)$–$1S_{3/2}(h)$, is calculated to be at ~298 nm, in agreement with our observation (~293 nm).

The size of the NCs can also be estimated from the excitonic profile. Table I compares the sizes obtained from the HRTEM and XRD measurements, as well as the calculated values based on effective mass approximation[12] and realistic tight binding calculation.[13] In addition, the narrow size distribution of the NCs is confirmed by the sharp optical absorption edge and well-defined excitonic feature. Broadening of the excitonic transition is primarily due to the inhomogeneity arising from size dispersion. By the use of Gaussian fitting,[14] a size distribution of ~ 11% can be estimated with the 240-meV width of the excitonic transition of $1S(e)$–$1S_{3/2}(h)$ observed in Fig. 2, consistent with the HRTEM analysis (~ 12%).

Figure 2 also displays a broad PL emission at 350–550 nm for the ZnS NCs. The PL emission peak (~ 440 nm), which is red-shifted compared to the excitonic transition (~ 293 nm), is consistent with the observation by Qu *et al.*[15] and Sapra *et al.*[8] for un-doped ZnS NCs. The change of excitation wavelength only leads to the alteration in the intensity of the emission peak. Sapra *et al.* attributed this strong emission band to the carrier recombination of the defect states, which are mostly on the surface of the NCs due to sulfur vacancies. The PLE spectrum of the ZnS NCs (Fig. 2) gives two well-resolved excitation bands (centered at 283 nm and 306 nm, respectively).

The room-temperature 3PA of the ZnS NCs in aqueous solution of 1-cm optical path was investigated with standard Z-scan technique. The 1-mJ, 1 kHz, 120-fs laser pulses were generated by a Ti:Sapphire regenerative amplifier (Quantronix, Titan), which was seeded by an erbium-doped fiber laser (Quantronix, IMRA). The details of the Z-scan setup can be found in Ref. 16. For comparison, similar 3PA measurements were conducted on a 0.5-mm-thick cubic ZnS bulk crystal (Semiconductor Wafer, Inc.) with laser polarization perpendicular to its <111> axis.

All the Z scans reported here were performed with excitation irradiances below the damage threshold, which was determined to ~ 130 GW/cm$^2$ for the ZnS NC solution by a reported method.[17]

Figure 3 illustrates the open-aperture (OA) Z-scan curves for the ZnS NCs and the ZnS bulk crystal at different excitation irradiances ($I_{00}$), which is defined as the peak, on-axis irradiance at the focal point ($z$ = 0) within the sample. By employing an analytical method,[16] the 3PA coefficients, $\gamma$, for the ZnS NC solution and the ZnS bulk crystal are found to be 0.000045 and 0.0016 cm$^3$/GW$^2$, respectively. The 3PA coefficients are related to the imaginary part of the fifth-order susceptibilities by $\gamma = 5\pi \,\mathrm{Im}\, \chi^{(5)} / \left(\lambda n_0{}^3 c^2 \varepsilon_0{}^2\right)$, where $n_0$ is the linear refractive index, $c$ the speed of light in vacuum, and $\varepsilon_0$ the dielectric constant in vacuum. Thus, the intrinsic 3PA coefficient of the NCs, $\gamma_{NC}$, can be derived as $\gamma_{NC} = \gamma_{solution} n_{0solution}^3 /(n_{0NC}^3 f_v |f|^6)$, where $f_v$ is the volume fraction of the NCs in the aqueous solution, and $f$ the local field correction that depends on the dielectric constant of the solvent and the NCs. The value of $f$ is ~0.58 while $f_v$ (~0.89%) can be accurately determined by elemental analysis. The intrinsic 3PA coefficient obtained for the ZnS NCs is 0.024 cm$^3$/GW$^2$, which is ~15 times larger than that of the bulk ZnS. This enhancement can be attributed to the quantum confinement effect[18] in ZnS NCs since our NCs' average radius (~ 1.3 nm) is much smaller than the Bohr exciton radius (~ 2.2 nm). Note that our measured 3PA coefficient for bulk ZnS is in good agreement with our recent report.[16] We can convert the 3PA coefficient of 0.000045 cm$^3$/GW$^2$ into the 3PA cross-section ($\sigma_3$) by the definition of $\sigma_3 = \frac{(\hbar\omega)^2 \gamma}{N_0}$, where $\hbar\omega$ is the photon energy, and $N_0$ the density of ZnS NCs in the solution. By using $N_0 = 1.1 \times 10^{18}$ cm$^{-3}$, we find $\sigma_3$ to be $2.7 \times 10^{-78}$ $cm^6s^2$ $photon^{-2}$ for the ZnS NCs, which is at least two orders of magnitude larger than that of ZnS bulk crystal.[19] Furthermore, this 3PA cross-section is four to five orders of magnitude higher than those of common UV fluorescent dyes,[20] and nearly one order of magnitude larger than that of CdS NCs,[5] which is one of the fluorescent semiconductor NCs for multiphoton-excited fluorescence imaging. The comparison of CdS NCs, ZnS NCs and ZnS bulk crystal is presented in Table I. Note that the 2PA cross-sections for ZnS NCs ($\sim 2.0 \times 10^{-46}$ $cm^4s$ $photon^{-1}$) are about one order of magnitude higher than that of CdS NCs ($\sim 10^{-47}$ $cm^4s$ $photon^{-1}$).[5,9]

As shown in Fig. 2, there exist surface (or defect) states below the lowest excitonic transition. These states could mediate multi-step excitation processes, which might lead to apparently high value of the 3PA cross section. To assess the magnitude of the 2PA, we plot $\mathrm{Ln}(1-T_{OA})$ vs. $\mathrm{Ln}(I_0)$, as detailed in Ref. 16. By the use of linear fit to the plots of $\mathrm{Ln}(1-T_{OA})$ vs. $\mathrm{Ln}(I_0)$, one can find the gradient to 2 for 3PA and 1 for 2PA. As shown in the insets of Fig. 3, the slope obtained ($s = 1.9$) confirms the dominance of 3PA in the ZnS NCs. In addition, we can also determine the 2PA and 3PA coefficients unambiguously with the use of a Z-scan theory recently developed for materials that possess 2PA and 3PA simultaneously.[21] Details of the calculation is not presented here but it verifies again that the 2PA is negligible in the ZnS NCs.

In the pump-probe experiments, we employed a cross-polarized, pump-probe configuration[16] with the same laser system used for the Z scans. The intensity ratio of the pump to the probe was kept at least 40: 1. Fig. 4 illustrates the degenerate transient transmission signals ($-\Delta T$) as a function of the delay time. For the bulk crystal, the transient transmission signals are mainly dominated by the autocorrelation function of the pump and probe pulses, which reveal that the 3PA plays a key role in the observed nonlinear absorption since 3PA is an instantaneous nonlinear process. When the excitation pump irradiance is increased to ~ 200 GW/cm$^2$, there is a long absorption tail with a characteristic time of ~ 100 ps or longer. This slow recovery process can be attributed to absorption of 3PA-excited free carriers in the bulk ZnS since the amplitude of the absorption tail grows proportionally to the cube of the excitation pump irradiance (not shown in Fig. 4). The oscillatory behavior observed at higher excitation irradiance might be attributed to saturation of 3PA or excitation of phonon mode.[22] However, both slow recovery and oscillatory behavior do not manifest themselves in the ZnS NCs with the excitation irradiance up to ~130 GW/cm$^2$, which is the photo-induced damage threshold. The main peaks in the measured dynamics seem to be broader in the NCs than in the bulk crystal. This may be attributed to group velocity dispersion since the ZnS-NC solution is contained in a 1-cm-thick quartz cell, while the thickness of the ZnS bulk crystal is 0.5 mm.

In summary, our study shows that the ZnS NCs possess a larger 3PA cross section than its bulk counterpart; and it is also nearly one order of magnitude greater than CdS NCs. More importantly, such a large 3PA is observed at 780 nm, an optimal wavelength of commercial Ti:sapphire femtosecond lasers.

**Table I.** Crystallite size, one-photon absorption, 2PA and 3PA of water-soluble ZnS NCs

| | HRTEM (nm) | XRD (nm) | Calculated size (nm) | | Lowest one-photon absorption transition (nm) | $\gamma_{NC}$ ($cm^3/GW^2$) | 3PA cross-section ($\sigma_3$) ($cm^6s^2$ $photon^{-2}$) | 2PA cross-section ($\sigma_2$) ($cm^4s$ $photon^{-1}$) |
|---|---|---|---|---|---|---|---|---|
| | | | NEMA[a] | FP-LAPW[b] | | | | |
| ZnS NCs | 2.5 | 2.2 | 2.5 | 2.6 | 293 | 0.024 | $2.7\times10^{-78}$ | $\sim 2.0\times10^{-46}$ [c] |
| Bulk ZnS | | | | | 337 | 0.0016 | $0.6\times10^{-80}$ [d] | |
| CdS NCs | | | | 3.9 | 440[e] | | $\sim 10^{-79}$ [e] | $\sim 10^{-47}$ [e] |

[a]Reference 12.

[b]Reference 13.

[c]Reference 9

[d]Reference 19.

[e]Reference 5.

**Figure Captions**

Fig. 1 (a) HRTEM micrograph and (b) XRD pattern (thick solid line :**—**) of the ZnS NCs. Gaussian fit of the data curve (thin solid line :—), and the individual deconvoluted peaks (dotted lines :···) are also included in (b).

Fig. 2 UV-vis absorption (solid line :—), PL (dashed line :---) and PLE (dotted line :···) spectra of the ZnS NCs. The excitation wavelength of PL spectrum is 290 nm, while the emission wavelength of PLE spectrum is 450 nm.

Fig. 3 Open-aperture Z-scans at different excitation irradiances ($I_{00}$) at 780 nm for (a) the ZnS NCs and (b) the ZnS bulk crystal. The symbols denote the experimental data while the solid lines are theoretically fitted curves. The insets show the plots of $\ln(1-T_{OA})$ vs. $\mathrm{Ln}(I_0)$; the solid lines represent the linear fits to the data.

Fig. 4 Degenerate, transient transmission measurements on the ZnS NCs at 780 nm at different excitation pump irradiances ($I_{00}$). The inset shows similar measurements on the ZnS bulk for comparison.

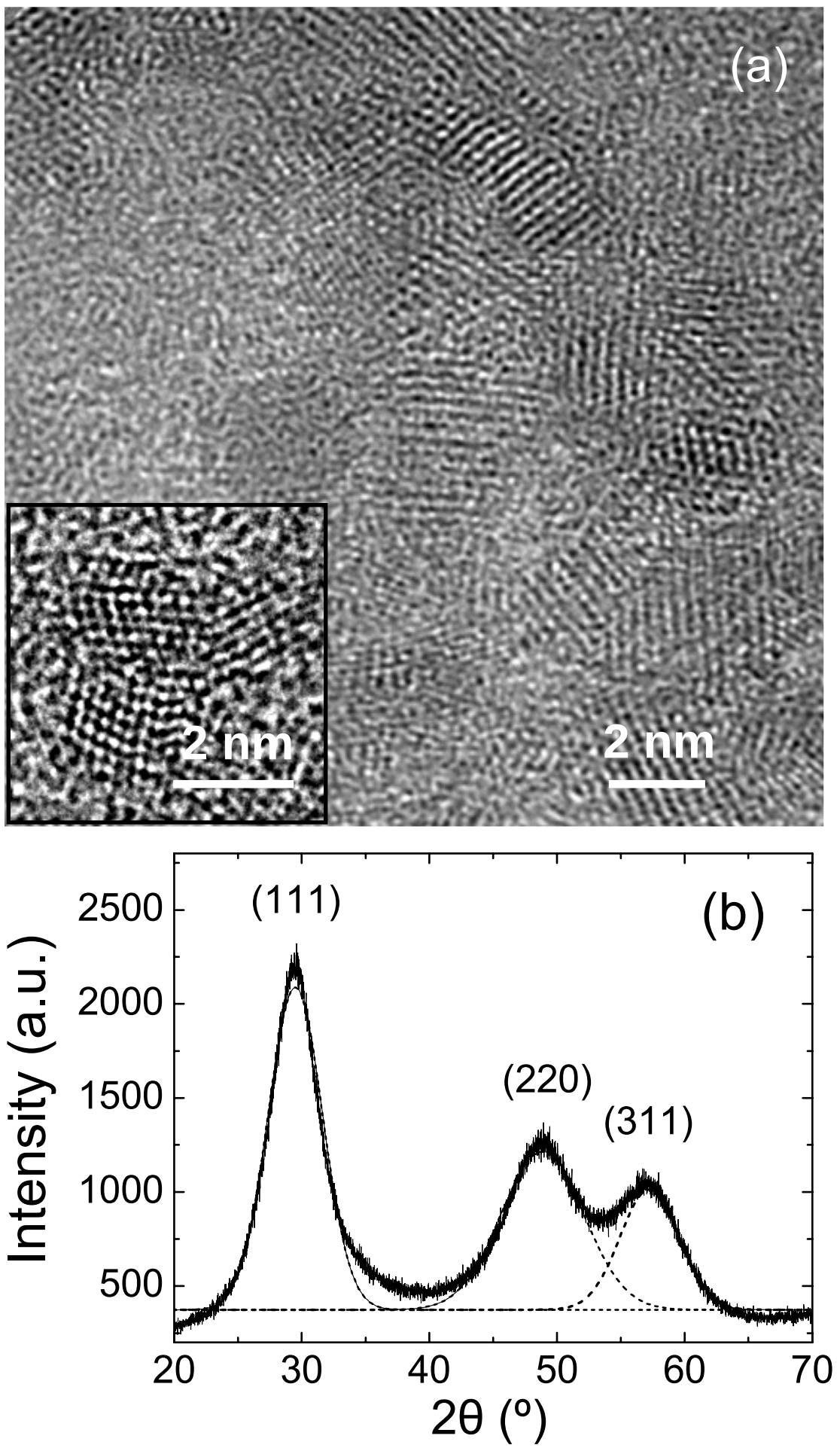

**Fig. 1.** **J. He *et al.***

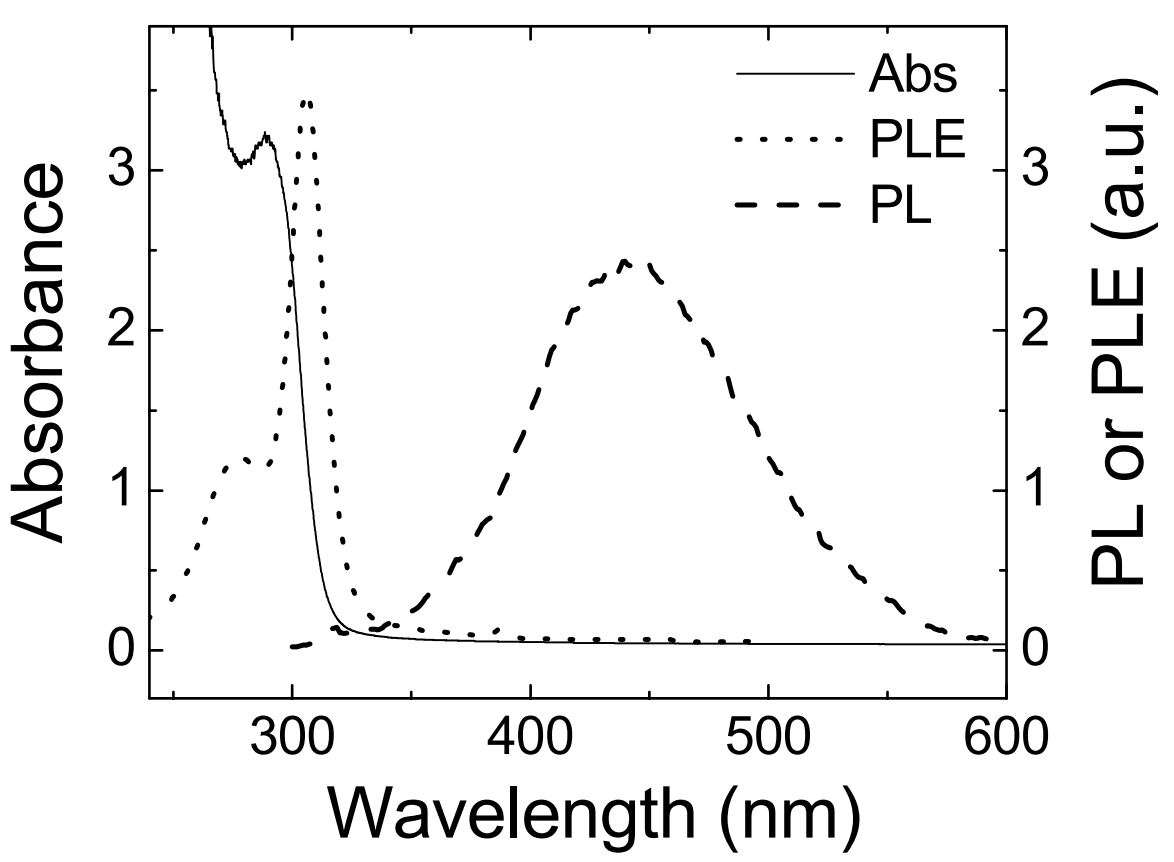

**Fig. 2. J. He *et al.***

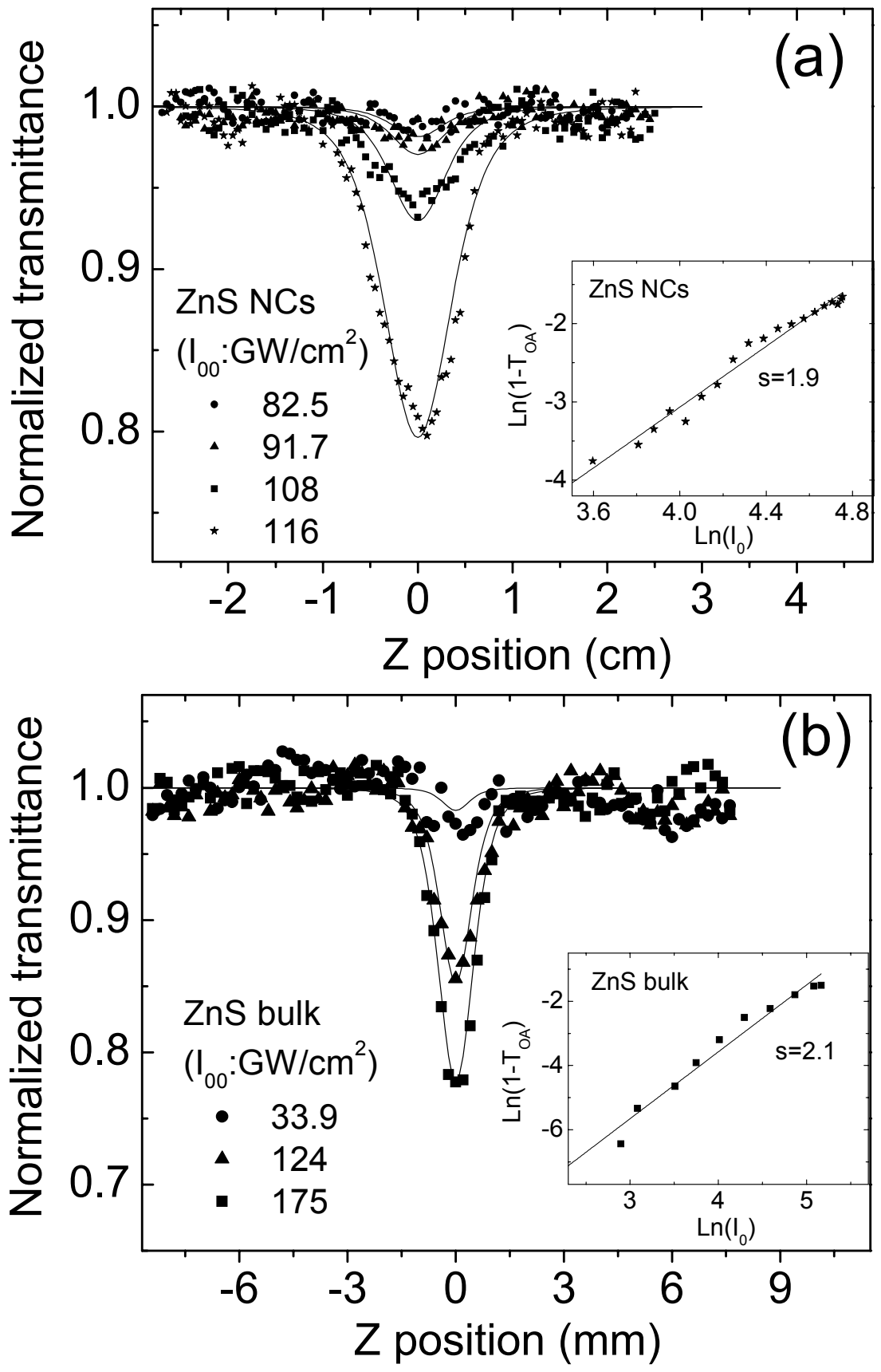

**Fig. 3. J. He *et al.***

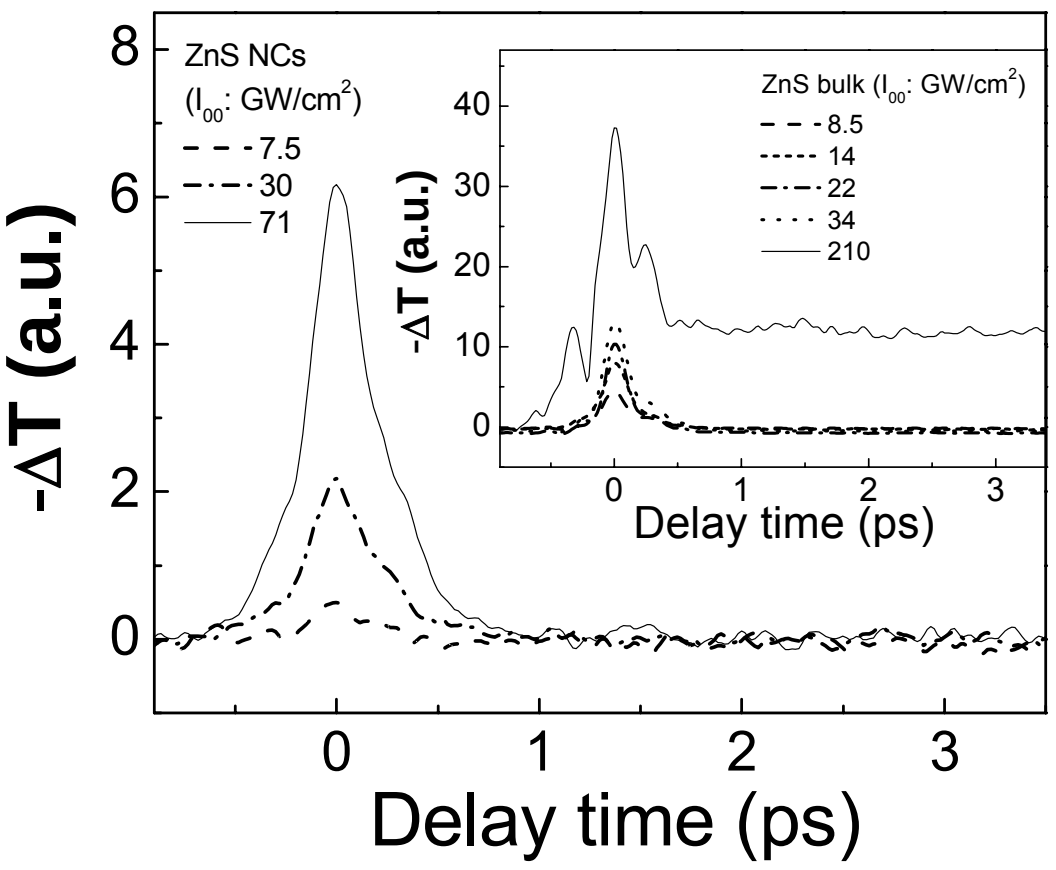

**Fig. 4.** **J. He *et al.***